\documentclass[useAMS,usenatbib]{mn2e}
\usepackage{epsf,amsfonts,amsmath,amssymb}
\usepackage{graphicx}
\usepackage{mathrsfs}
\usepackage{color}
\usepackage[usenames,dvipsnames,svgnames,table]{xcolor}
\usepackage{array}
\usepackage{ulem}
\usepackage{rotating}
\usepackage{pdflscape}
\usepackage{multirow}

\newcolumntype{C}[1]{>{\centering\let\newline\\\arraybackslash\hspace{0pt}}m{#1}}
\def\xr{X-ray~}

\title[GRB prompt efficiencies]{A Revised Analysis of Gamma Ray Bursts' prompt efficiencies}

\author[P. Beniamini et al.]{Paz Beniamini
\thanks{E-mail:paz.beniamini@gmail.com}, Lara Nava and Tsvi Piran\\
Racah Institute for Physics, The Hebrew University, Jerusalem, 91904, Israel\\
}

\begin{document}

\date{Accepted ... Received ...; in original form ...}

\pagerange{\pageref{firstpage}--\pageref{lastpage}} \pubyear{2002}
\maketitle

\label{firstpage}

\begin{abstract}
The prompt Gamma-Ray Bursts' (GRBs) efficiency is an important clue on the emission mechanism producing the $\gamma$-rays. Previous estimates of the kinetic energy of the blast waves, based on the X-ray afterglow luminosity $L_X$, suggested that this efficiency is large, with values above 90\% in some cases. This poses a problem to emission mechanisms and in particular to the internal shocks model. These estimates are based, however, on the assumption that the X-ray emitting electrons are fast cooling and that their Inverse Compton (IC) losses are negligible. The observed correlations between $L_X$ (and hence the blast wave energy) and $E_{\gamma\rm ,iso}$, the isotropic equivalent energy in the prompt emission, has been considered as observational evidence supporting this analysis. It is reasonable that the prompt gamma-ray energy and the blast wave kinetic energy are correlated and the observed correlation corroborates, therefore, the notion $L_X$ is indeed a valid proxy for the latter. Recent findings suggest that the magnetic field in the afterglow shocks is significantly weaker than was earlier thought and its equipartition fraction, $\epsilon_B$, could be as low as $10^{-4}$ or even lower. Motivated by these findings we reconsider the problem, taking now IC cooling into account. We find that the observed $L_X-E_{\gamma\rm ,iso}$ correlation is recovered also when IC losses are significant. For small $\epsilon_B$ values the blast wave must be more energetic and we find that the corresponding prompt efficiency is significantly smaller than previously thought. For example, for $\epsilon_B\sim10^{-4}$ we infer a typical prompt efficiency of $\sim15\%$. 
\end{abstract}


\section{Introduction}
\label{Int}
Gamma-Ray Bursts (GRBs) are extremely energetic pulses of $\gamma$-rays, associated with a relativistic jet launched following the core collapse of a massive star or a compact binary merger.
Energy dissipation internal to the jet is thought to be responsible for the emission of the prompt $\gamma$-rays, while the subsequent collision between the jet and the  external environment produces the  longer-lived afterglow.

Two critical quantities in this model are the energy radiated in the first prompt phase, and the energy that remains in the blast-wave and that powers the afterglow.
While the first can be directly estimated from prompt observations, the latter can  be inferred  only indirectly from afterglow observations.
The sum gives the total amount of initial explosion energy, an important information that constrains the nature of the progenitor.
The ratio indicates the efficiency of the prompt phase (i.e. the efficiency of the dissipation mechanism times the efficiency of the radiative process).

In models involving hydrodynamic jets, large dissipation efficiencies are unlikely realized: maximal values are estimated to be $\lesssim 0.2$ \citep{Kobayashi(1997), Daigne(1998), Lazzati(1999), Kumar(1999),Beloborodov(2000), Guetta(2001),Beniamini(2013),Vurm(2013)}.
In electromagnetic jets, it may be possible to obtain higher dissipative efficiencies \citep[see e.g.][]{Zhang(2011)}. However the situation is much less certain (see e.g. \citealt{Granot(2015),Kumar(2015),BG(2016)}).
The efficiency is unlikely to approach unity: magnetic field lines approaching the reconnection zone are unlikely to be exactly anti-parallel, and a significant portion of the EM energy could remain undissipated.
Furthermore, a major challenge in models that rely on synchrotron to produce the prompt radiation is to explain the
observed spectral indices below the sub-MeV peak. This may be viable if the electrons are only ``marginally fast cooling" or if their spectra is modified by IC cooling.
Both possibilities suggest that the efficiency of radiation is only moderate \citep{Derishev(2001),Bosnjak(2009),Daigne(2011),Beniamini(2013),Beniamini(2014)}.
Determining the overall efficiency would therefore give important clues on the still uncertain nature of the mechanism responsible for the prompt radiation. 
It follows that inferring reliable estimates of the (isotropically equivalent) kinetic energy $E_{\rm kin}$ that remains in the blast wave after the prompt phase is of paramount importance.

Under certain conditions, $E_{\rm kin}$ can be quite firmly estimated from afterglow observations.
If observations are performed at a frequency where the emission is dominated by fast cooling electrons (i.e. a frequency larger than the characteristic synchrotron frequencies) {\it and } if these electrons do not suffer from significant Inverse Compton (IC) losses,
then the afterglow luminosity at such a frequency provides a robust estimate of the energy stored in the accelerated electrons,
which is in turn directly related to the kinetic energy of the blast wave \citep{Kumar(2000),Freedman(2001)}.

It has been argued that electrons emitting X-ray afterglow radiation fulfil these conditions.
A correlation between the (isotropically equivalent) X-ray luminosity $L_X$ and the (isotropically equivalent) energy released during the prompt phase $E_{\gamma\rm ,iso}$ has indeed been observed in both long and short GRBs. This supported the notion that the \xr luminosity
is a good proxy for the kinetic energy, and hence it must  be produced by fast cooling electrons that undergo negligible IC losses.
Under this assumption, several studies have exploited X-ray observations to estimate the energies of GRB blast waves and eventually also the prompt efficiencies $\epsilon_\gamma$ \citep{Freedman(2001),Berger(2003), LRZ(2004), Berger(2007),Nysewander(2009),D'Avanzo(2012),Wygoda(2015)}.
Most of these studies inferred relatively low kinetic energies that correspond to large prompt efficiencies 
$\epsilon_\gamma=E_{\gamma\rm ,iso}/(E_{\gamma\rm ,iso}+E_{\rm kin})$.
Values larger than $50\%$ and up to more than $90\%$ have been estimated in some cases \citep{Granot(2006),Ioka(2006),Nousek(2006),Zhang(2007)}. 
The discovery of \xr plateaus in many X-ray light-curves has increased the severity of the efficiency problem and 
poses even more serious  challenge  for the internal shocks model. 

Recently, the location of the cooling frequency compared to the X-ray frequency and the relevance of IC losses have been brought into question. 
In a study involving bursts with temporally extended GeV emission, \cite{Beniamini(2015)} have shown (with multi wavelength modelling performed under the assumption that GeV radiation originated at the external shocks) that \xr emitting electrons are either slow cooling or they suffer from significant IC losses,
making the \xr flux not directly related to the blast wave energy. In this scenario, high-energy (GeV) radiation has been proposed to be a better proxy for the kinetic energy,
since it is always deep in the fast cooling regime and it is  less affected by IC losses (due to Klein-Nishina suppression).
The tight correlation  found between the luminosity of the temporally extended GeV emission and $E_{\gamma\rm ,iso}$\citep{Nava(2014)}
supports this scenario.
If this is the case, however, a question immediately arises: how can there be a correlation between $L_X$ and $E_{\gamma\rm ,iso}$ if the \xr luminosity is not a proxy for the blast wave energy content $E_{\rm kin}$?

Both slow cooling and significant IC losses arise in low magnetic field regions, i.e. for small values of the magnetic equipartition parameter, $\epsilon_B\lesssim 10^{-2}$. 
Such small values are required for GeV-detected bursts if this emission arises from external shocks  \citep{KBD(2009),KBD(2010),Lemoine(2013),Beniamini(2015)}.
Moreover, several recent studies not based on GRBs detected at GeV energies have found similar results, with an inferred $\epsilon_B$ distribution that peaks around $10^{-4}$ and extends down to $10^{-6}-10^{-7}$ (\citealt{RBD(2014),Santana(2014), Zhang(2015),Wang(2015)}).
A theoretical explanation for such small values in the context of a decaying turbulence has been provided by \cite{Lemoine(2013)} and \cite{Lemoine(2013b)}.
These recent findings suggest another urgent question: how do the estimates of the kinetic energies (and in turn the estimates of the prompt efficiencies) change if the assumption on the typical values of $\epsilon_B$ in the range $0.1-0.01$ are modified, and more precisely, if smaller values are considered.

In this paper we address these two main issues. We explore whether the observed correlation between \xr luminosities and prompt energetics implies that $L_X$ is a proxy for the blast wave energy and can be used as a tool to derive the prompt efficiency.  We then examine  how are the estimates of these two quantities  affected by different choices of $\epsilon_B$.
We proceed as follows. First we characterize the observed correlation using a sample of Swift GRBs (section~\ref{sec:obs}). 
Then we consider the standard synchrotron/synchrotron self-Compton (SSC) afterglow model and derive (for different assumptions on the typical values and distributions of all the free parameters) the expected $L_X-E_{\gamma\rm ,iso}$ correlation and compare it with the observations. 
For those sets of parameters for which the slope, normalization, and scatter of the observed correlation are reproduced, we check what is the cooling regime of the electrons emitting X-rays, and the relevance of their SSC losses. 
We find that the observed correlation can be reproduced also when SSC cooling is not negligible.
To understand the origin of this result we present both simplified analytic estimates (section~\ref{sec:model}) and  detailed numerical results (section~\ref{sec:sim}).
We also  use the simulated \xr luminosities to derive the blast wave kinetic energies and prompt efficiencies under the assumption of fast cooling and negligible IC cooling, as usually done with real X-ray observations. We compare these derived quantities with the simulated ones, to infer by how much the derived values differ from the simulated ones. The conclusions are discussed and summarized in section~\ref{sec:conclusions}.

\section{Observations}
\label{sec:obs}
\begin{figure*}
\includegraphics[scale=0.3]{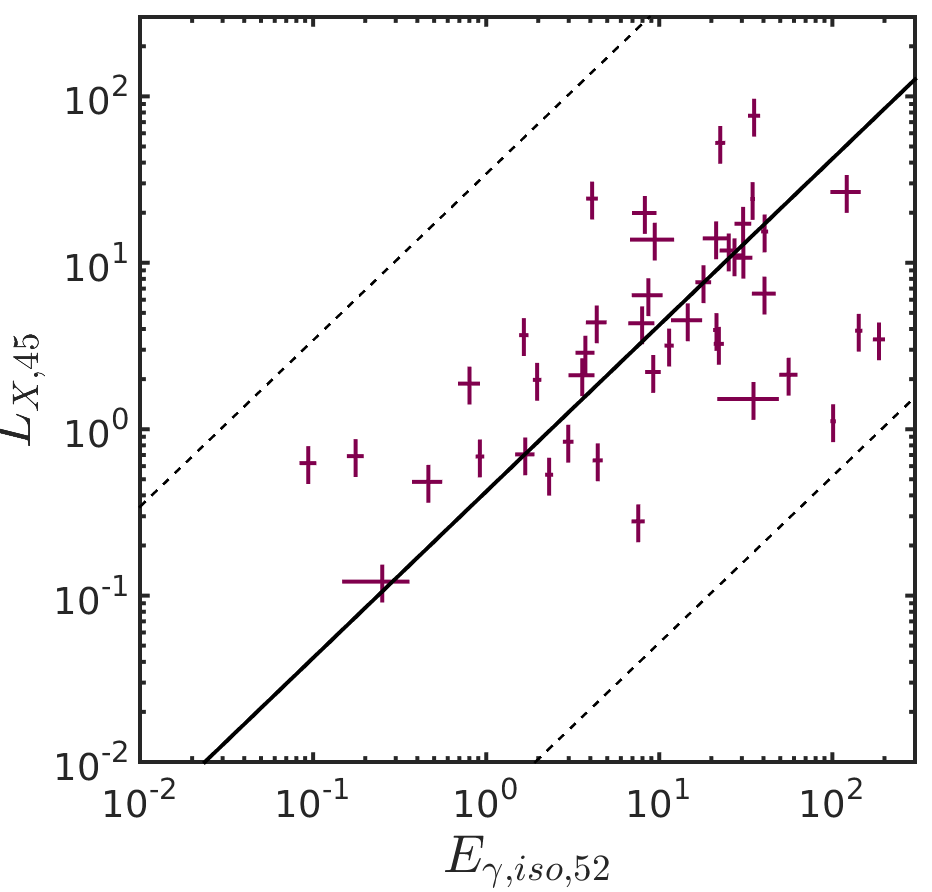}
\includegraphics[scale=0.3]{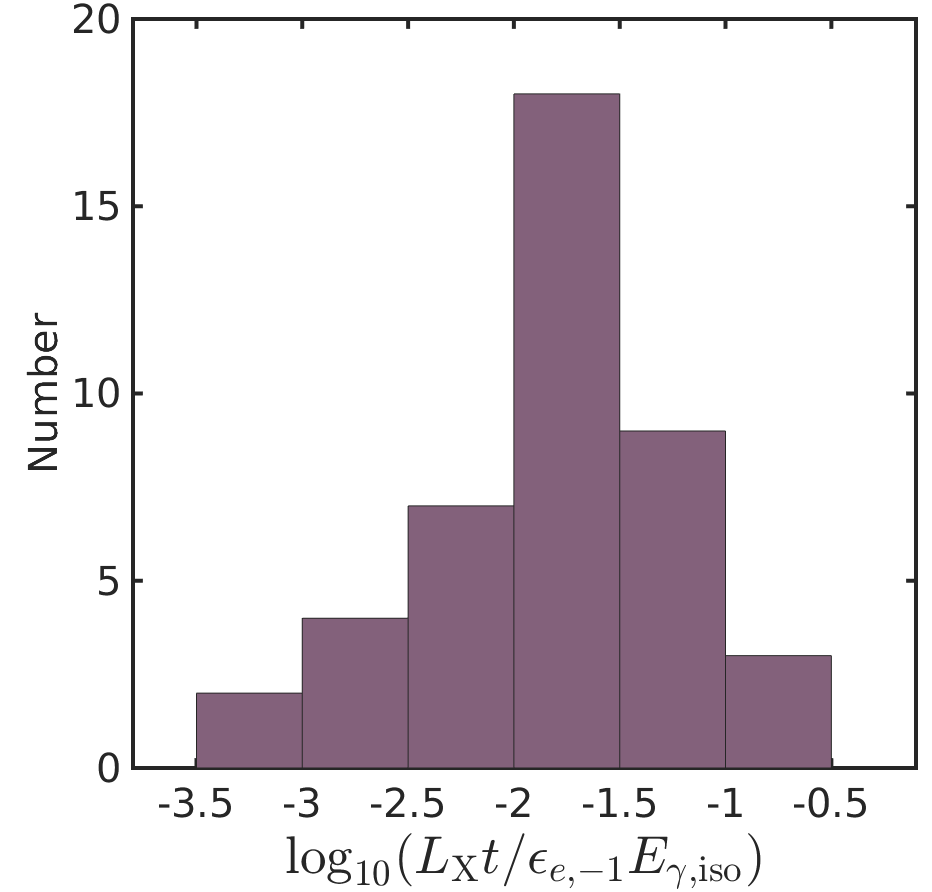}
\caption{Correlation between the afterglow X-ray luminosity $L_X$ and the prompt energy $E_{\gamma\rm ,iso}$ for the sample of bursts presented in \citet{D'Avanzo(2012)}. $L_X$ is measured 11 hours after the burst trigger, integrated in the energy range 2-$10\,$keV.
In the left panel, the solid line depicts the best linear fit, corresponding to $L_{X,45}=0.42~E_{\gamma\rm ,iso,52}$, while dashed lines show the $3 \sigma$ scatter. In the right panel, following \citet{Nakar(2007)}, we plot a histogram of $L_X t/(\epsilon_{e,-1} E_{\gamma,\rm iso})$.
If the X-ray flux is produced by synchrotron
from fast cooling electrons with negligible IC, and the fraction of energy stored in the electrons is $\epsilon_e\approx 0.1$, this ratio would provide an
estimate for $E_{\rm kin}/E_{\gamma,\rm iso})$.
} 
\label{fig:observations}
\end{figure*}

In order to compare the results of our simulations with observations we need to select a sample of GRBs with measured $L_X$ and $E_{\gamma\rm ,iso}$.
We use the so-called BAT6 sample, a sample of long Swift GRBs carefully selected to be almost complete in redshift (for details see \citealt{Salvaterra(2012)}).
The necessary information is available for 43 events.
For this sample, the correlations $L_X-E_{\gamma\rm ,iso}$ (for four different choices of the rest frame time at which $L_X$ is computed) are presented in \cite{D'Avanzo(2012)}.
In the following we  consider $L_X$ at 11 hours: this is the most common value used in this and other correlation studies
and it allows us a comparison of results derived using different samples.
For the BAT6 sample, the values of $L_X$ (integrated in the rest frame energy range 2-$10\,$keV) can be found in Table 1 of \cite{D'Avanzo(2012)}, while the values of the prompt energy $E_{\gamma\rm ,iso}$ are reported in \cite{Nava(2012)}. 
The resulting correlation is shown in Fig.~\ref{fig:observations}. 
The best linear fit between $L_X$ and $E_{\gamma\rm ,iso}$ is given by:
\begin{equation}
\label{eq:correl}
L_{X,45}=0.42\,E_{\gamma\rm ,iso,52}\quad \rm with \quad\sigma_{log (L_{X}/E_{\gamma\rm ,iso})}=0.64~
\end{equation}
where $\sigma_{log (L_{X}/E_{\gamma\rm ,iso})}$ is the 1$\sigma$ scatter (measured in log-log space)
 \footnote{We use here and elsewhere in the text the notation $Q_x=Q/10^x$ in c.g.s. units as well as base 10 logarithms.}.
The correlation between  $L_X$ at 11 hours and $E_{\gamma \rm ,iso}$ has been investigated by different authors using different samples (see \citealt{Nysewander(2009)}, \citealt{Margutti(2013)}, and \citealt{Wygoda(2015)} for recent investigations).
These studies find statistically significant correlations between $L_X$ and $E_{\gamma\rm ,iso}$.
The slope, normalization and scatter of the correlations discussed in these other studies are consistent with the one found in the BAT6 sample.
Based on these findings, it has been argued that $L_X$ must be a good proxy for the kinetic blast wave energy $E_{\rm kin}$.

\section{Analytic estimates}
\label{sec:model}
According to the standard forward shock afterglow theory  if the X-ray emitting electrons are fast cooling then the X-ray luminosity, $L_X $ is tightly related to the kinetic energy in the blast wave  as $E_{\rm kin} /(1+Y)$, where $Y$ is the Compton parameter.
Previous studies \citep[e.g.][]{Kumar(2000),Freedman(2001),Berger(2003), LRZ(2004), Berger(2007),Nysewander(2009),D'Avanzo(2012),Wygoda(2015)} assumed that Compton losses are small and neglected the factor $1+Y$. 
These estimates obtained  low values of $E_{\rm kin}$ and hence implied 
puzzling large values of the prompt efficiency $\epsilon_\gamma$.
As $E_{\rm kin}$ is related to $E_{\gamma\rm ,iso}$ the observed correlation between $L_X$ and $E_{\gamma\rm ,iso}$ 
has  
been interpreted as supporting the validity of the overall analysis and in particular the assumption of negligible Compton losses. 
We show here that the correlation persists even  when Compton losses are important and $Y\gg 1$.  In this case the inffered 
prompt efficiencies are much lower.

We begin by considering, once more,  the model with no IC losses.
In this case the X-ray luminosity (integrated in the rest frame energy range 2-$10\,$keV), is given by:

\begin{equation} 
\label{eq:standard}
L_{X,45}\!=\!
\left\{ \!
  \begin{array}{l}
 1.6\ f(p) \bigg(\frac{1+z}{2}\bigg)^{2+p\over 4}\ \bar\epsilon_{e,-1}^{\, \,p-1}\ \epsilon_{B,-4}^{p-2\over 4} \\ \ E_{\gamma\rm ,iso,52}^{2+p\over 4}\ \bigg(\frac{1-\epsilon_{\gamma}}{\epsilon_{\gamma}}\bigg)^{2+p\over 4}
 \ t_{\rm 11 hours}^{2-3p\over 4} \quad \mbox{for ISM}\\
 \\
 1.7\ g(p) \bigg(\frac{1+z}{2}\bigg)^{2+p\over 4}\ \bar\epsilon_{e,-1}^{\, \,p-1}\ \epsilon_{B,-4}^{p-2\over 4} \\ \ E_{\gamma\rm ,iso,52}^{2+p\over 4}\ \bigg(\frac{1-\epsilon_{\gamma}}{\epsilon_{\gamma}}\bigg)^{2+p\over 4}
 \ t_{\rm 11 hours}^{2-3p\over 4}\quad \mbox{for wind}\\
  \end{array} \right.
\end{equation}
where $p$ is the power-law index of the electrons' energy spectrum, 
$f(p)$ and $g(p)$ are dimensionless functions of order unity defined such that $f(p=2.2)=g(p=2.2)=1$,
$\epsilon_e$ is the fraction of shock dissipated energy gained by electrons ($\bar \epsilon_e\equiv \epsilon_e (p-2)/(p-1)$), $t_{\rm 11 hours}$ is the time since burst and $z$ is the cosmological redshift.
For typical values of $p$, Eq.~\ref{eq:standard} can be approximated by $L_{X,45}(11\,\rm h) \approx 2\times \, \epsilon_{e,-1} E_{kin,52} \, \approx 2\, \epsilon_{e,-1} [( {1-\epsilon_\gamma})/{\epsilon_\gamma}] E_{\gamma\rm ,iso,52}$, leaving out here a weak extra dependence on $\epsilon_B$.

Eq.~\ref{eq:standard} has been traditionally used to infer $E_{\rm kin}$ from $L_X$. A comparison of $E_{\rm kin}$ with $E_{\gamma\rm ,iso}$ can be used to estimate the prompt efficiency.
The observed normalization of the $ L_X-E_{\gamma\rm ,iso}$ correlation (see Eq.~\ref{eq:correl}) implied a 
large average efficiency, $\epsilon_{\gamma}\approx 0.8$. 
According to Eq. \ref{eq:standard}, $[L_{X,45}/E_{\gamma\rm ,iso,52}]_{\mbox{\tiny{no IC}}} \propto \epsilon_e\epsilon_{\gamma}^{-1}$. 
To account for the observed correlation the dispersion in  both $\epsilon_{\gamma}$ and $\epsilon_e$ must be relatively small.
As one is a prompt quantity while the other is an afterglow quantity, there is no a priori reason to expect the two to be correlated. The observed  spread in the correlation (see Fig. \ref{fig:observations}) limits, therefore, the variability of each one of those quantities to about 1 dex (see a discussion in \citealt{Nava(2014)}).

Although the assumption of negligible IC losses is unclear for the X-ray emitting electrons, it must hold for the GeV emitting electrons 
for which IC is deep in the Klein Nishina region (see \citealt{Beniamini(2015)} for a discussion). 
If the GeV luminosity  is indeed produced by synchrotron in the forward shock, it  should then be correlated to $E_{\gamma\rm ,iso}$ according to Eq.~\ref{eq:standard}. A correlation consistent with this scenario (but at an earlier observed time) has been indeed found by \cite{Nava(2014)}.

We take now  into account IC losses by the X-ray emitting electrons.
We assume in this section that the IC cooling is in the Thomson regime and
using the synchrotron forward shock model we estimate the 
\xr afterglow luminosity $L_X$ as a function of the afterglow free parameters.
We assume that $\nu_m<\nu_c$ (where $\nu_m$ is the synchrotron frequencies), and $2<p<3$.
The Compton parameter $Y$ is given by \citep{Sari(2001)}:
\begin{equation}
\label{Y1}
 Y=\frac{\epsilon_e}{\epsilon_B (3-p) (1+Y)}\bigg(\frac{\nu_m}{\nu_c}\bigg)^{p-2\over 2}~.
 \end{equation}
Since we are interested in the situation where IC cooling is important, we explore the behaviour for $Y\gg1$ (in our numerical estimates we will not be limited to this regime).
In this limit:
\begin{equation}
\label{Y2}
Y\approx 
\left\{
  \begin{array}{l}
21\ \hat{f}(p)\  E_{\rm kin,53}^{p-2 \over 2(4-p)}\ n_0^{p-2 \over 2(4-p)}\ \epsilon_{B,-4}^{p-3 \over 4-p}\times\ \\
\times\epsilon_{e,-1}^{p-1 \over 4-p}\ t_{\rm 11 hours}^{2-p \over 2(4-p)}\ \bigg(\frac{1+z}{2}\bigg)^{p-2 \over 2(4-p)}\quad \mbox{ISM}\\
\\
25\ \hat{g}(p)~\!   A_*^{p-2 \over 4-p}\ \epsilon_{B,-4}^{p-3 \over 4-p}\ \epsilon_{e,-1}^{p-1 \over 4-p}\ t_{\rm 11 hours}^{2-p \over 4-p} \bigg(\frac{1+z}{2}\bigg)^{p-2 \over 4-p}\quad \! \mbox{Wind}\\
  \end{array} \right.
\end{equation}
where $\hat{f}(p)$ and $\hat{g}(p)$ are dimensionless functions of order unity defined such that $f(p=2.2)=g(p=2.2)=1$, $n_0$ is the particle density in $\mbox{cm}^{-3}$ and $A_*\equiv A/(5\times10^{11}\mbox{g/cm})$ is the wind parameter.
We have normalized $\epsilon_e,\epsilon_B,n_0,A_*$ to the values implied by recent literature \citep{KBD(2009),KBD(2010),Lemoine(2013),RBD(2014),Santana(2014),Zhang(2015),Wang(2015)}.

In the regime $Y\gg1$, at a given fixed time, $Y$ depends very weakly on the unknown kinetic energy and density (see Eq. \ref{Y2}), and somewhat more strongly on the fraction of energy stored in the electrons and magnetic field:
\begin{itemize}
\item $Y\propto E_{\rm kin}^{1/18}n^{1/18}\epsilon_B^{-4/9} \epsilon_e^{2/3}$ for $p=2.2$ and ISM medium
\item $Y\propto E_{\rm kin}^{1/6}n^{1/6}\epsilon_B^{-1/3}\epsilon_e$ for $p=2.5$ and ISM medium
\item $Y\propto A_*^{1/9}\epsilon_B^{-4/9}\epsilon_e^{2/3}$ for $p=2.2$ and wind medium
\item $Y\propto A_*^{1/3}\epsilon_B^{-1/3}\epsilon_e$ for $p=2.5$ and wind medium.
\end{itemize}
At this stage we already see that the dispersion that would be introduced due to the $Y$ parameter is relatively small.
The implied dispersion will become even weaker when we go back to $L_X$.

To determine the X-ray luminosity, one has first to determine the cooling regime of the  X-ray producing electrons, i.e. the location of the cooling frequency $\nu_c$ as compared to the X-ray frequency $\nu_x$.
Following \cite{Granot(2002)} and introducing a multiplicative factor of $(1+Y)^{-2}$ to account for IC cooling, we obtain (as long as $Y \gg 1$):
\begin{equation}
\label{nuc2}
\nu_c\approx 
\left\{
  \begin{array}{l}
6.3\times 10^{15}\ \tilde{f}(p)\  E_{\rm kin,53}^{-p \over 2(4-p)}\ n_0^{-2 \over 4-p}\ \epsilon_{B,-4}^{-p \over 2(4-p)}\times \\
\times\epsilon_{e,-1}^{-2(p-1) \over 4-p}\ t_{\rm 11hours}^{3p-8 \over 2(4-p)}\ \bigg(\frac{1+z}{2}\bigg)^{-p \over 2(4-p)}\mbox{Hz}\quad \mbox{ISM}\\
\\
1.7\times 10^{14}\ \tilde{g}(p)\  E_{\rm kin,53}^{1 \over 2}\ A_*^{-4 \over 4-p}\ \epsilon_{B,-4}^{-p \over 2(4-p)}\times \\
\times\epsilon_{e,-1}^{-2(p-1) \over 4-p}\ t_{\rm 11hours}^{3p-4 \over 2(4-p)}\ \bigg(\frac{1+z}{2}\bigg)^{-(4+p) \over 2(4-p)}\mbox{Hz}\quad \mbox{Wind}\\
  \end{array} \right.
\end{equation}
where $\tilde{f}(p),\tilde{g}(p)$ are dimensionless functions such that $\tilde{f}(p=2.2)=\tilde{g}(p=2.2)=1$. 
According to these simple estimates, at $\sim 11$ hours, unless both $\epsilon_B$ and $n$ are very small
$\nu_c< \nu_x$, i.e. X-ray radiation at this time is typically emitted by ``fast cooling" electrons. 
The first condition for using X-ray luminosities as a tool to derive $E_{\rm kin}$ appears then to be satisfied in most cases.
It still remains to be seen whether the  $L_X-E_{\gamma\rm ,iso}$ correlation is expected  in the regime $Y \gg 1$ (where the flux above $\nu_c$
is significantly suppressed by SSC cooling) and under what conditions it matches the observed one.

To derive $L_X$ we divide  the expression for the specific flux at frequencies larger than $\nu_c$  \citep{Granot(2002)} by a factor (1+$Y$). 
We then integrate the specific flux between $2\,$keV and $10\,$keV to get the luminosity.
We obtain: 
\begin{equation} 
\label{eq:EgammatoLumX}
\frac{L_{X,45}}{E_{\gamma\rm ,iso,52}}\!=\!
\left\{ \!
  \begin{array}{l}
0.84\ \bar{f}(p)\  \epsilon_{\gamma,-1}^{{(p-2)^2 \over 4(4-p)}-1}E_{\gamma\rm ,iso,52}^{-(p-2)^2 \over 4(4-p)}\ n_0^{2-p \over 2(4-p)}\ \epsilon_{B,-4}^{-(p^2-2p-4) \over 4(4-p)}\\
\times\epsilon_{e,-1}^{(p-1)(3-p) \over 4-p}\ t_{\rm 11hours}^{3p^2-12p+4 \over 4(4-p)}\ \bigg(\frac{1+z}{2}\bigg)^{-(p^2-12) \over 4(4-p)}\quad \mbox{for ISM}\\
\\
0.74\ \bar{g}(p)\  \epsilon_{\gamma,-1}^{{2-p \over 4}-1}E_{\gamma\rm ,iso,52}^{p-2 \over 4}\ A_*^{2-p \over 4-p}\ \epsilon_{B,-4}^{-(p^2-2p-4) \over 4(4-p)} \\
\times\epsilon_{e,-1}^{(p-1)(3-p) \over 4-p}\ t_{\rm 11hours}^{p(3p-10) \over 4(4-p)}\ \bigg(\frac{1+z}{2}\bigg)^{-(p^2+2p-16) \over 4(4-p)}\mbox{for Wind}\\
  \end{array} \right.
\end{equation}
where  $\bar{f}(p),\bar{g}(p)$ are dimensionless functions such that $\bar{f}(p=2.2)=\bar{g}(p=2.2)=1$.

The relation between $L_X$ and $E_{\gamma\rm ,iso}$ is almost linear, as the ratio $L_X/E_{\gamma\rm ,iso}$ depends only weakly on $E_{\gamma\rm ,iso}$: 
$L_X/E_{\gamma\rm ,iso}\propto E_{\gamma\rm ,iso}^{-0.018}$ for $p=2.2$ in an ISM medium ($L_X/E_{\gamma\rm ,iso}\propto E_{\gamma\rm ,iso}^{0.05}$ for wind) and
$L_X/E_{\gamma\rm ,iso}\propto E_{\gamma\rm ,iso}^{-0.09}$ for $p=2.5$ in an ISM medium ($L_X/E_{\gamma\rm ,iso}\propto E_{\gamma\rm ,iso}^{0.125}$ for wind).

The ratio $L_{X,45}/E_{\gamma\rm ,iso,52}$ depends very weakly on the density, and approximately scales as: $[L_{X,45}/E_{\gamma\rm ,iso,52}]_{\mbox{\tiny{with IC}}} \propto \epsilon_B^{1/2} \epsilon_e^{1/2}\epsilon_{\gamma}^{-1}$.
This results should be compared with the situation of fast cooling without IC suppression: $[L_{X,45}/E_{\gamma\rm ,iso,52}]_{\mbox{\tiny{no IC}}} \propto \epsilon_e\epsilon_{\gamma}^{-1}$.
The scaling in $\epsilon_{\gamma}$ is the same. Clearly, no correlation will appear in either case if the prompt efficiency varied significantly from one  burst to another. 
{When IC losses are negligible, the scatter of the correlation is related to the scatter of the parameters by $\sigma^2_{Log(L_X/E_{\gamma \rm,iso})}=\sigma^2_{\log\epsilon_e}+\sigma^2_{\log\epsilon_\gamma}$,
where, following the reasoning at the top of the section, we have assumed that $\epsilon_e$ and $\epsilon_{\gamma}$ are independent.
With significant IC cooling  $\sigma^2_{Log(L_X/E_{\gamma \rm,iso})}=0.25\sigma^2_{\log\epsilon_e}+0.25\sigma^2_{\log\epsilon_B}+0.5\sigma_{\log \epsilon_e \epsilon_B}+\sigma^2_{\log\epsilon_\gamma}$,
where $\sigma_{\log \epsilon_e \epsilon_B}$ is the correlation coefficient between $\log_{10}(\epsilon_e)$ and $\log_{10}(\epsilon_B)$.
Depending on the conditions determined by the forward shock, $\sigma_{\log \epsilon_e \epsilon_B}$ may be either positive or negative. 
The additional scatter due to the new parameter $\epsilon_B$ is compensated by 
a weaker dependence on $\epsilon_e$. Since both are microphysical parameters of the afterglow shock a possible anti correlation between the two 
can even reduce the overall scatter.


Keeping $\epsilon_e$ fixed, we note that the observed value 0.42 of the normalisation (see Eq.~\ref{eq:correl}) can be reproduced by playing with the values of $\epsilon_B$ and $\epsilon_\gamma$:
a reasonable efficiency ($\epsilon_{\gamma}\approx 0.15$) is recovered for $\epsilon_B=10^{-4}$, while higher values of $\epsilon_B$ require higher values of $\epsilon_\gamma$ (as $\epsilon_B$ increases,
the assumption $Y \gg1$ breaks down and we cannot use the equations derived in this section any more).
We demonstrated that even for $Y\gg1$ a correlation with the correct slope and normalization is expected. 

Large $Y$ might imply a bright SSC component at GeV energies, detectable with the Fermi/LAT. 
At $\sim11\,$hours, under the most conservative assumption that the entire energy stored in
the electrons is emitted as IC radiation we estimate a SSC flux $\sim 2\times 10^{-12}E_{\rm kin,53}\epsilon_{e,-1} t_{\rm 11hours}^{-1}d_{L28}^{-2}\mbox{ergs~}\mbox{cm}^{-2}\mbox{sec}^{-1}$. 
This is orders of magnitude weaker than detectability limits with Fermi/LAT in the
$>0.1\,$GeV range, which are typically $10^{-8}\mbox{ergs~}\mbox{cm}^{-2}\mbox{sec}^{-1}$ \citep{ACK13}, and at best may approach $10^{-9}\mbox{ergs~}\mbox{cm}^{-2}\mbox{sec}^{-1}$, \citep[see e.g.][]{ACK12}.
Moreover, the IC peak is expected to reside at energies $>10\,$GeV.
This would reduce the prospects of detectability even further, since the LAT effective area quickly decreases at large energies.
At earlier times ($t\sim10-10^2\,$seconds), and for the most energetic bursts (with $E_{\rm kin}\gtrsim 10^{54}$ergs) we can expect a total flux of $\sim 2\times 10^{-8}\mbox{ergs~}\mbox{cm}^{-2}\mbox{sec}^{-1}$.
Even though marginally detectable, this SSC component might explain ($t\sim10-10^2\,$seconds) photons with energies that exceed the energy limit of synchrotron radiation  \citep{Wang(2013),Tang(2014)}.

\section{Numerical simulations}
\label{sec:sim}
Motivated by the approximate analytical scalings found in \S \ref{sec:model} we examine here 
numerically  under which conditions the slope, normalization and scatter of the correlation can be reproduced.
We consider synchrotron radiation from a forward shock afterglow, including IC corrections to the synchrotron spectrum.
While in the previous section we discussed results in the two extreme regimes $Y\gg1$ and $Y<1$, here we solve numerically eq.~\ref{Y1}, which is valid in both regimes.
We find that the observed correlation is reproduced for a wide range of typical values and dispersions in the distributions of the afterglow parameters, also when SSC cooling is relevant.

We have calculated, first,  for different values of  $\epsilon_B$ and $n$ what is the value of $\epsilon_\gamma$ needed in order to recover the normalization of the $L_X-E_{\gamma\rm ,iso}$ correlation, for both ISM and wind external media.
 Fig. \ref{fig:normalization}, depicts the results for $\epsilon_e=0.1$ and $p=2.2$ (the results depend only weakly on $p$).
In both cases (ISM  and wind) the resulting efficiency depends weakly on $n$ (with an exception at low values of the wind parameter that we discuss later). 
The value of $\epsilon_\gamma$ depends strongly on the assumed value of $\epsilon_B$: for large values of $\epsilon_B$, SSC cooling is negligible, eq.~\ref{eq:standard} can be used, and a relatively large value of $\epsilon_\gamma$ is inferred. 
For smaller values of $\epsilon_B$, larger kinetic energies are needed in the outflow and hence lower values of the efficiency are found. For relatively low values of the density and $\epsilon_B$ (low-left corner of the plane in Fig.~\ref{fig:normalization}), the X-ray emitting electrons are in the slow cooling regime (see Eq. \ref{nuc2}).
In this regime, only a fraction of the electrons' energy is actually emitted as radiation (be it synchrotron or IC).  The required prompt efficiency $\epsilon_{\gamma}$ decreases as the density decreases, as more kinetic energy is needed in the outflow when the system gets deeper into the slow cooling regime.

\begin{figure*}
\includegraphics[scale=0.45]{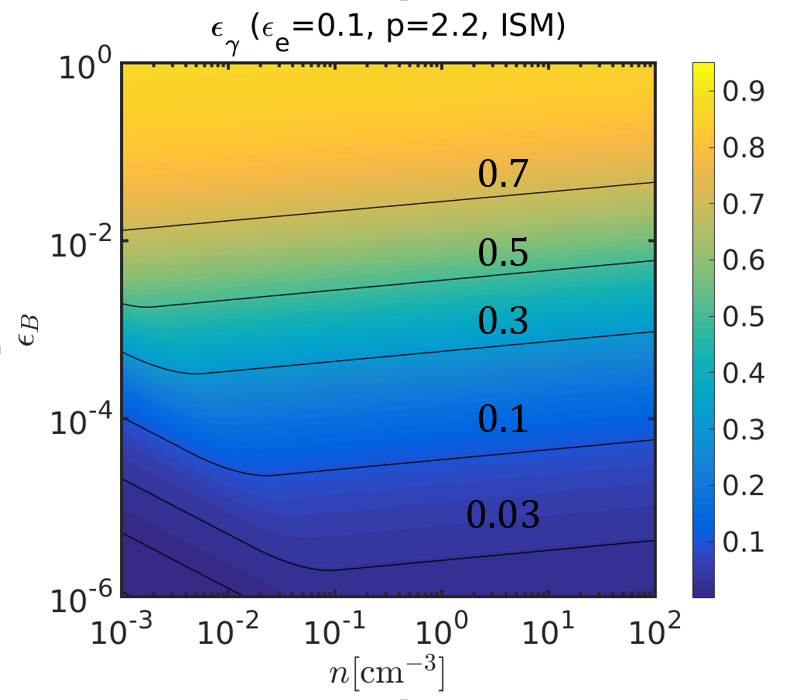}
\includegraphics[scale=0.45]{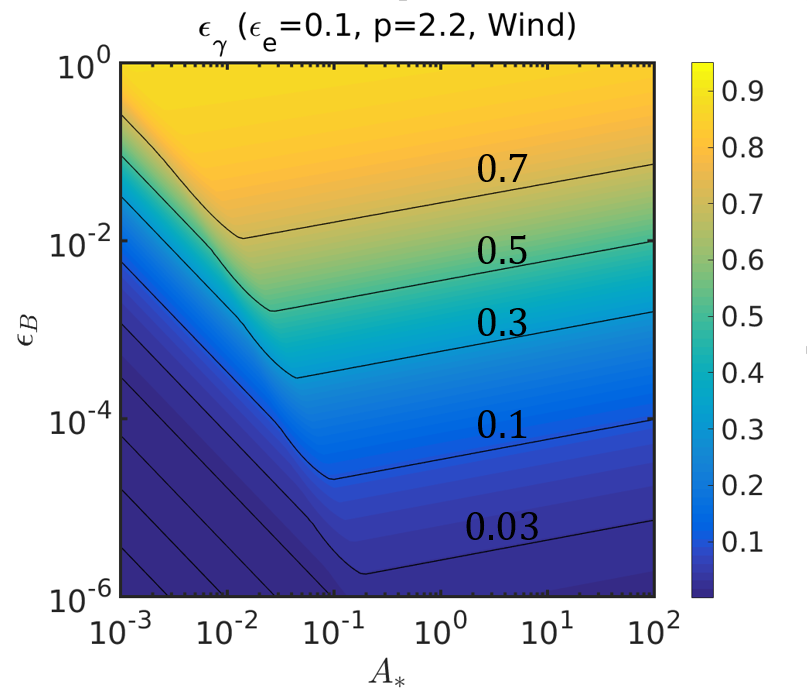}\\
\includegraphics[scale=0.45]{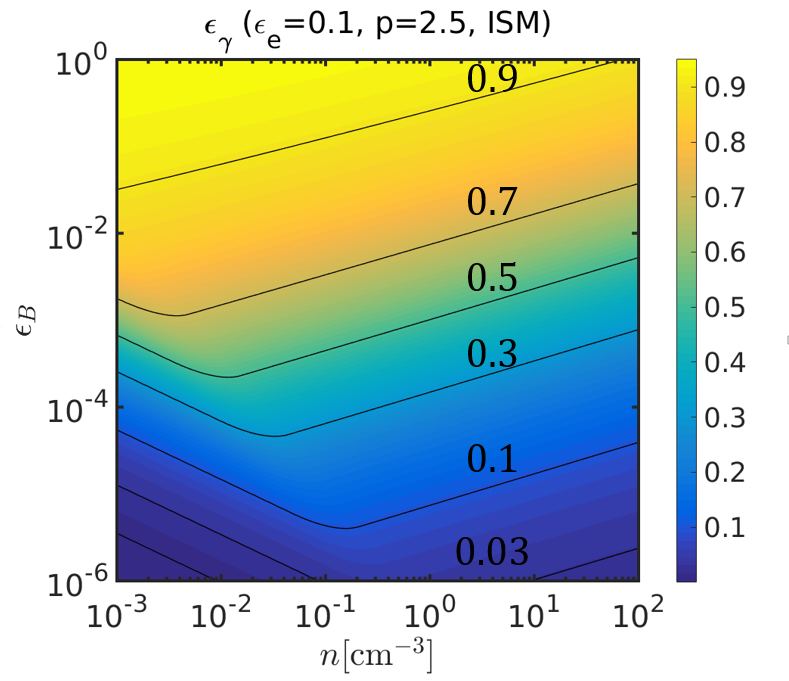}
\includegraphics[scale=0.45]{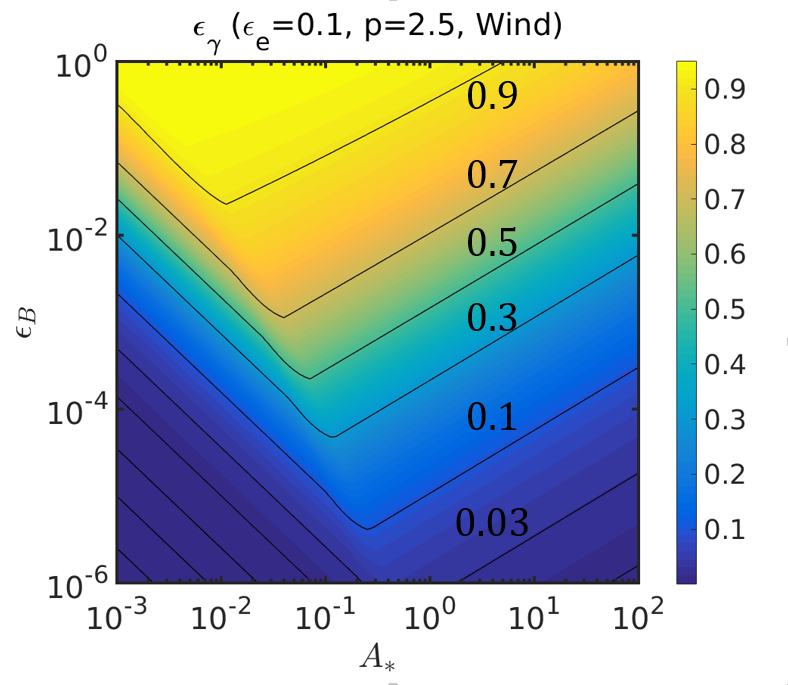}
\caption{$\epsilon_{\gamma}$ implied by  the normalization of the observed $L_{X}-E_{\gamma\rm ,iso}$ correlation as a function of the density 
and $\epsilon_B$ for $\epsilon_e=0.1$ (left panels: ISM;  right panels: wind; top panels: $p=2.2$; bottom panels: $p=2.5$).}
\label{fig:normalization}
\end{figure*}

The scatter of the $L_{X}-E_{\gamma\rm ,iso}$ correlation depends on the width of the distributions of the parameters involved.
The fact that a correlation is observed with a given dispersion  limits  the dispersion of such parameters.
In order to estimate the widths of the relevant distributions we apply a Monte Carlo method: we assign a given distribution to each free parameter,
randomly draw a value and using the forward shock afterglow synchrotron + IC model we calculate $L_X$ for each realization.
We draw $10^5$ realizations and compare the obtained correlation with the observed one and derive the conditions required to reproduce the observations.

For two of the parameters ($E_{\gamma\rm ,iso}$ and $z$) the distributions are deduced from observations.
In order to compare the simulated correlation with the observed correlation in the \cite{D'Avanzo(2012)} BAT6 sample
we use that sample to obtain the distributions of $E_{\gamma\rm ,iso}$ and $z$. The distribution of $E_{\gamma\rm ,iso}$ is taken from observations of bursts with known redshift. 
Using \cite{D'Avanzo(2012)} we consider a log-normal distribution with a mean value: $\langle E_{\gamma\rm ,iso}\rangle=8\times 10^{52}$ergs and a standard deviation $\sigma_{log E_{\gamma\rm ,iso}}=0.75$.
For redshifts, we fit the distribution of bursts used by \cite{D'Avanzo(2012)} and take a log-normal distribution with a peak at $z=1$ and a standard deviation of $0.3$ dex.
For the other parameters we consider lognormal distributions for $\epsilon_B$, $n$ (or $A_*$), and $\epsilon_e$, and either a fixed value or a uniform distribution for $p$.
For $\epsilon_e$ we choose $\langle \log_{10}(\epsilon_e)\rangle =-1$ and $\sigma_{log \epsilon_e}=0.3$ in all the simulations: $\epsilon_e$ is indeed confined both from observations \citep{Santana(2014),Nava(2014)} and from
numerical simulations \citep{Sironi(2011)}  to have a narrow distribution peaked around $\epsilon_e=0.1$  (see \citealt{Beniamini(2015)} for a detailed discussion).
For $\epsilon_B$ and $n$ we test different average values and widths. 
The intrinsic distributions of these parameters are less certain. However, typical values for the 1 $\sigma$ dispersion
found for both these parameters in GRB modelling are of order 1 dex \citep{Santana(2014),Zhang(2015)}. Therefore, these are the canonical values that we consider here. Since \cite{Soderberg(2006)} find a somewhat wider distribution for $n$ (consistent with $\sigma=1.7$ dex), we explore also the possibility
of wider density distributions ($\sigma=1.5,2$ dex). Since $L_X$ depends very weakly on $n$, its dispersion can be significantly increased with minor effects to the overall results.
Finally, $\epsilon_{\gamma}$ and its scatter are chosen such that the normalization and scatter of the $L_{X}-E_{\gamma\rm ,iso}$ correlation are reproduced (see Eq.~\ref{eq:correl}).
Considering the detectability limits of Swift/XRT \citep{Gehrels(2004)}, we apply a lower limit on the X-ray flux of $\sim 2\times10^{-14}\mbox{ergs} \mbox{ cm}^{-2} \rm s^{-1}$.

\def\arraystretch{1.5}
\begin{table*}
\begin{center}
\begin{tabular}{| c | c | c | c | c || c | c | c | c |}
\hline
 & \multicolumn{4}{c||}{Simulations' input parameters} & \multicolumn{4}{c|}{Results} \\
\hline
\multicolumn{1}{|C{1.6cm}|}{\centering varying \\ parameter} 
& Medium 
& $p$ 
& \multicolumn{1}{C{1.8cm}|}{\centering $ \log_{10}(\epsilon_B)$ \\ $\pm\sigma_{\log_{10}(\epsilon_B)}$} 
&  \multicolumn{1}{C{1.8cm}||}{\centering $ \log_{10}(n)$ \\ $\log_{10}(A_*)$} 
& $ \epsilon_{\gamma,\rm real}$& $\epsilon_{\gamma,\rm 2}$
& $\frac{E_{\rm kin,2}}{E_{\rm kin,real}}$ 
&  \multicolumn{1}{C{1.cm}|}{\centering$\%$ fast \\ cooling} \\

\hline
\multirow{6}{*}{$\epsilon_B$} & ISM & 2.2 & $-1\pm 1$ & $0\pm 1$ & $0.78_{-0.25}^{+0.22}$ & $0.75_{-0.19}^{+0.25}$  & $0.52_{-0.29}^{+0.65}$  & 99 \\ 
&ISM & 2.2 & $-2\pm 1$ & $0\pm 1$ & $0.61_{-0.2}^{+0.31}$ & $0.76_{-0.19}^{+0.24}$& $0.28_{-0.17}^{+0.45}$  & 97\\ 
&ISM & 2.2 & $-3\pm 1$ & $0\pm 1$ &  $0.36_{-0.13}^{+0.2}$ & $0.75_{-0.2}^{+0.25}$ & $0.11_{-0.07}^{+0.23}$& 96\\ 
&ISM & 2.2 & $-4\pm 1$ & $0\pm 1$  & $0.16_{-0.07}^{+0.12}$ & $0.74_{-0.21}^{+0.29}$ & $0.04_{-0.03}^{+0.08}$ & 92 \\  
&ISM & 2.2 & $-5\pm 1$ & $0\pm 1$ & $0.06_{-0.03}^{+0.05}$ & $0.74_{-0.2}^{+0.26}$ &  $0.01_{-0.008}^{+0.028}$  & 88 \\ 
&ISM & 2.2 & $-6\pm 1$ & $0\pm 1$ & $0.02_{-0.01}^{+0.02}$ & $0.73_{-0.22}^{+0.27}$ &  $0.004_{-0.003}^{+0.009}$ & 82 \\ 
\hline
\multirow{2}{*}{$n$}&ISM & 2.2 & $-4 \pm 1$ & $-1\pm 1$ & $0.17_{-0.06}^{+0.09}$ & $0.75_{-0.2}^{+0.27}$ & $0.04_{-0.03}^{+0.09}$ & 73 \\ 
&ISM & 2.2 & $-4 \pm 1$ & $1\pm 1$ & $0.14_{-0.06}^{+0.12}$ & $0.74_{-0.22}^{+0.26}$ & $0.03_{-0.02}^{+0.07}$  & 99 \\ 
&ISM & 2.2 & $-4 \pm 1$ & $0\pm 1.5$ & $0.16_{-0.06}^{+0.09}$ & $0.75_{-0.2}^{+0.25}$ & $0.03_{-0.02}^{+0.07}$  & 84 \\ 
&ISM & 2.2 & $-4 \pm 1$ & $0\pm 2$ & $0.16_{-0.05}^{+0.08}$ & $0.76_{-0.2}^{+0.24}$ & $0.03_{-0.02}^{+0.08}$  & 75 \\ 
\hline
\multirow{2}{*}{density}  &wind & 2.2 & $-2\pm 1$ & $0\pm 1$ & $0.61_{-0.18}^{+0.25}$ & $0.8_{-0.18}^{+0.2}$ & $0.24_{-0.15}^{+0.41}$ & 98\\ 
profile&wind & 2.2 & $-4\pm 1$ & $0\pm 1$ & $0.16_{-0.07}^{+0.12}$ & $0.76_{-0.21}^{+0.24}$ & $0.03_{-0.02}^{+0.08}$ & 89 \\
\hline
\multirow{2}{*}{$p$} &ISM & [2.1-2.7] & $-2 \pm 1$ & $0\pm 1$  &  $0.61_{-0.2}^{+0.31}$ & $0.87_{-0.12}^{+0.13}$ & $0.15_{-0.1}^{+0.3}$ & 96 \\ 
&ISM & [2.1-2.7] & $-4 \pm 1$ & $0\pm 1$ & $0.16_{-0.07}^{+0.12}$ & $0.82_{-0.18}^{+0.18}$ & $0.02_{-0.01}^{+0.05}$ & 88 \\ 
&ISM & 2.5 & $-2 \pm 1$ & $0\pm 1$  &  $0.72_{-0.21}^{+0.28}$ & $0.91_{-0.11}^{+0.09}$ & $0.11_{-0.07}^{+0.2}$ & 93 \\ 
&ISM & 2.5 & $-4 \pm 1$ & $0\pm 1$ & $0.26_{-0.11}^{+0.18}$ & $0.92_{-0.1}^{+0.08}$ & $0.02_{-0.01}^{+0.04}$ & 80 \\ 
\hline
\multirow{2}{*}{$\sigma_{\log \epsilon_B}, \sigma_{\log n}$} &ISM & 2.2 & $-2\pm 1.2$ & $0\pm 1.2$ & $0.61_{-0.15}^{+0.19}$ & $0.8_{-0.14}^{+0.2}$ & $0.25_{-0.17}^{+0.49}$ & 95 \\
&ISM & 2.2 & $-4\pm 1.2$  & $0\pm 1.2$ & $0.16_{-0.03}^{+0.04}$ & $0.73_{-0.22}^{+0.27}$ & $0.04_{-0.03}^{+0.1}$ & 89 \\
\hline
\end{tabular}
 \caption{List of the input parameters (on the left) and results (on the right) for different simulations. 
We fix all the afterglow parameters and vary one parameter at a time (as indicated in the first column). 
For the case of $p=[2.1-2.7]$, $p$ is drawn from a uniform distribution between 2.1 and 2.7.
For the results we report the allowed range (in order to fit the observed correlation and scatter) for the ``real'' prompt efficiency $\epsilon_{\gamma,real}$
(calculated using the simulated kinetic and $\gamma$-ray energies),
the prompt efficiency $\epsilon_{\gamma,2}$ as inferred from the simulated luminosities applying eq.~\ref{eq:standard},
the average ratio between the kinetic energy inferred from eq.~\ref{eq:standard}
and the input kinetic energy $E_{\rm kin,real}$, and the fraction of simulated GRBs for which X-rays are emitted
by electrons that are fast cooling. Reported errors are all at the 1$\sigma$ level.}
\label{tbl:dispersions}
\end{center}
\end{table*}

A summary of different input parameters for which $\sigma_{log \epsilon_B},\sigma_{log n}\geq 1$ and for which observations are satisfactorily reproduced is reported in Table~\ref{tbl:dispersions}.
This is of course not an exhaustive list, as the correlation could also be reproduced with a smaller scatter in
$\epsilon_B,n$ by considering a larger scatter in $\epsilon_{\gamma},\epsilon_e$.
As long as the dispersion in the intrinsic parameters satisfies $\sigma_{log \epsilon_B},\sigma_{log n}\leq1.2$ (or for instance $\sigma_{log \epsilon_B}\leq1$, $\sigma_{log n}\leq2$),
the correlation in the $L_{X}-E_{\gamma\rm ,iso}$ plane is recovered.
For each of the realizations, we also estimate, from the calculated \xr luminosities and the input $E_{\gamma\rm ,iso}$, 
the kinetic energy $E_{\rm kin,2}$ and efficiency $\epsilon_{\gamma,2}$ that would have been derived using eq.~\ref{eq:standard}, namely, assuming fast cooling and neglecting SSC.
We perform these estimates for an ISM medium, $\epsilon_e=0.1, \epsilon_B=0.01, n=1\mbox{cm}^{-3}$ and $p=2.2$ for all bursts.
Table~\ref{tbl:dispersions} summarizes $\epsilon_{\gamma,2}$, the ratio $E_{\rm kin,2}/E_{\rm kin,real}$
and the percentage of simulated GRBs for which $\nu_x>\nu_c$.
The results depend strongly on the assumed average value of $\epsilon_B$, and they depend very weakly on the mean value of $n$, on the nature of the external medium and on the value of $p$ (see Table \ref{tbl:dispersions}).
As expected, for low values of $\epsilon_B$ the values of the kinetic energy and efficiency derived assuming $Y\lesssim1$ and fast cooling
deviate significantly from those used for the simulations. Eq.~\ref{eq:standard} fails to recover the true (i.e. simulated) values of the parameters,
smaller kinetic energies are inferred and consequently, larger prompt efficiencies.

\begin{figure*}
\includegraphics[scale=0.28]{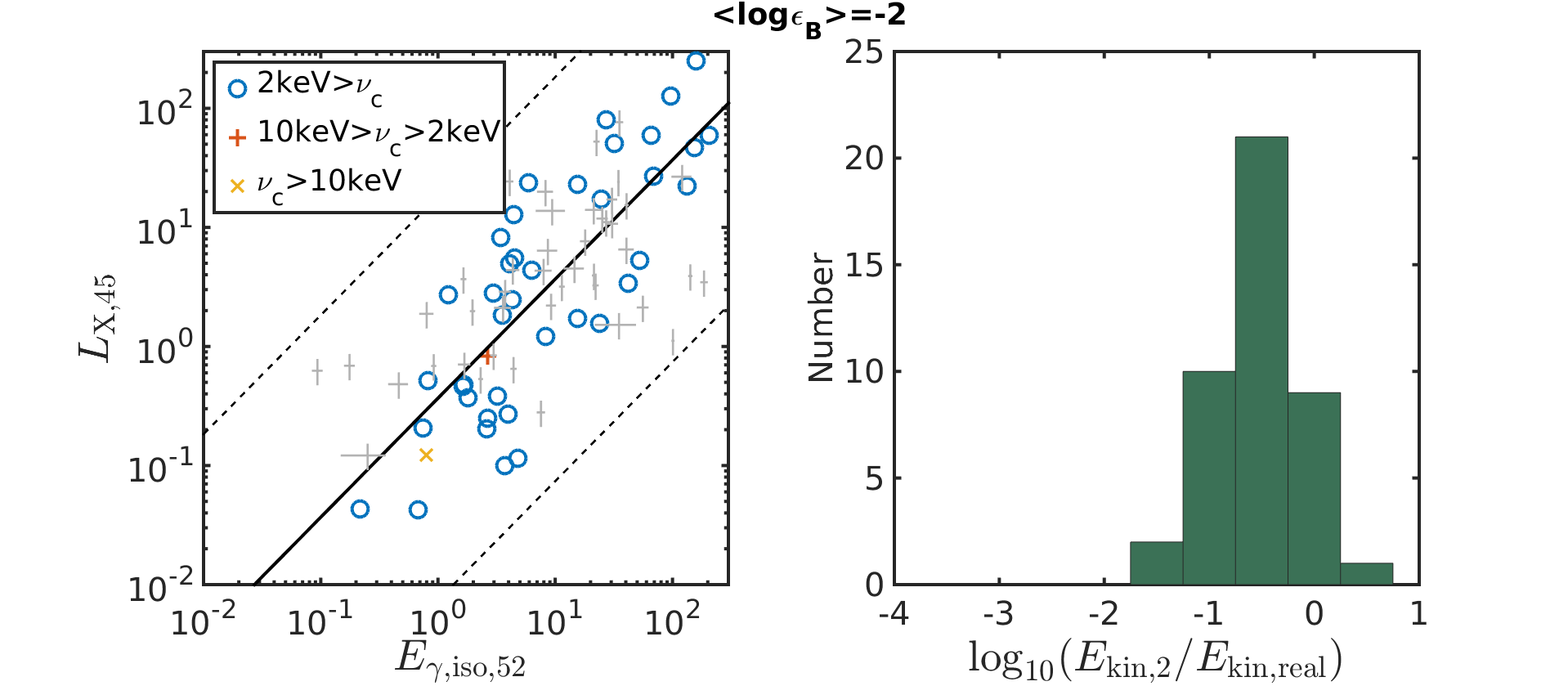}\\
\includegraphics[scale=0.28]{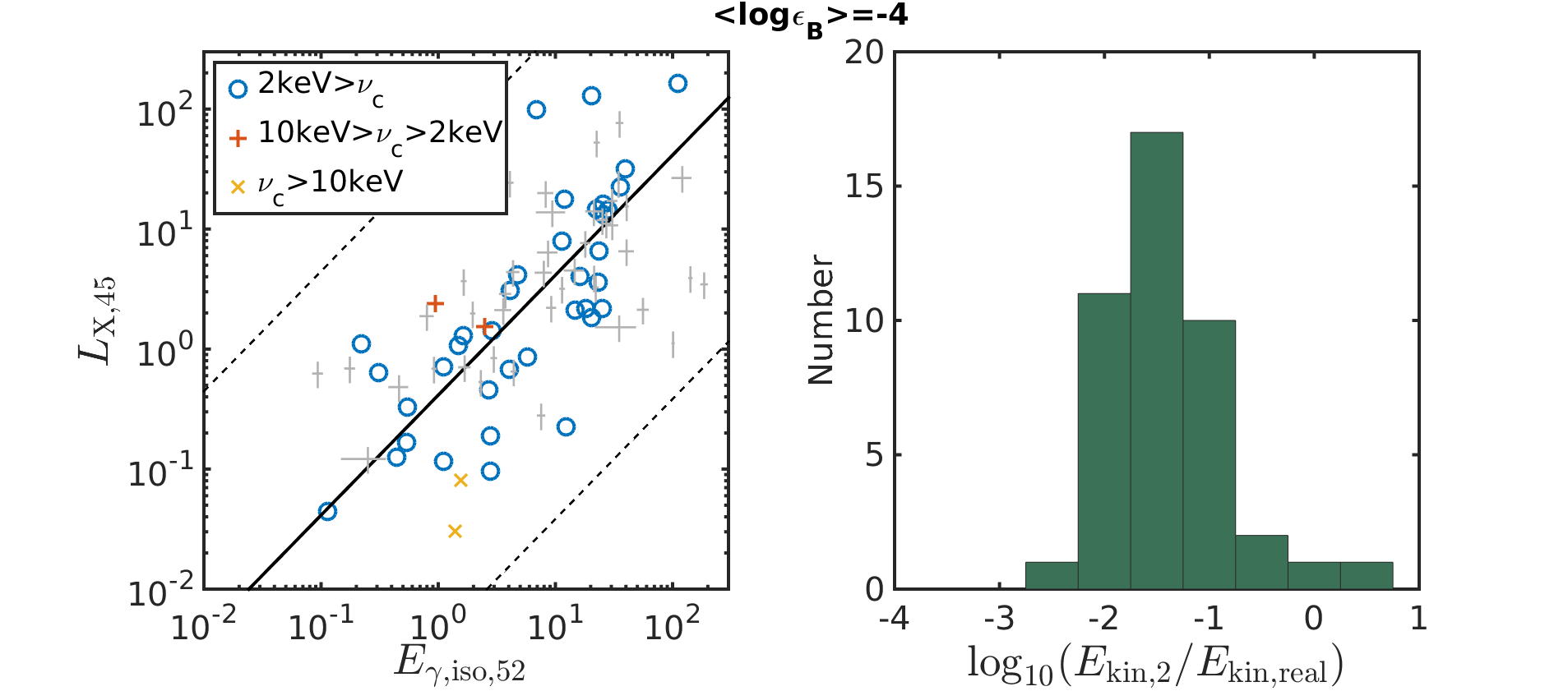}\\
\includegraphics[scale=0.28]{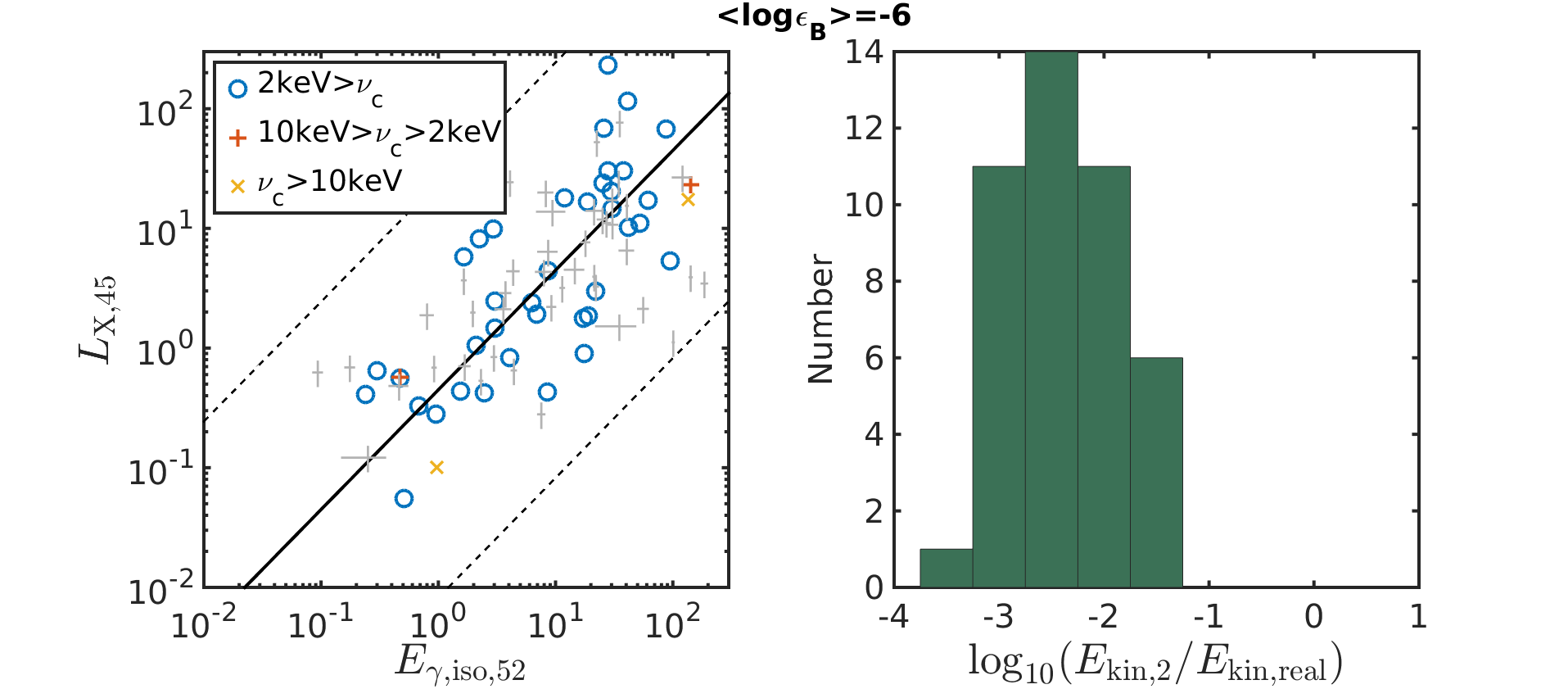}
\caption{Results of MC simulations for different assumptions on the median value of the magnetic field: $\epsilon_B=10^{-2}$ (upper panels), $\epsilon_B=10^{-4}$ 
(middle panels), $\epsilon_B=10^{-6}$ (lower panels). 
All the other parameters are the same in the three different simulations:  $\sigma_{log \epsilon_B}=1$,
an ISM medium with $\langle\log_{10}(n [\mbox{cm}^{-3}])\rangle=0, \sigma_{log n}=1$, a log-normal distribution of $\epsilon_e$ with $\langle \log_{10}
(\epsilon_e)\rangle =-1, \sigma_{log \epsilon_e}=0.3$, $p=2.2$, and a redshift distribution which is log-normal with a peak at $z=1$ and a standard deviation of $0.3$ 
dex.
$\epsilon_{\gamma}$ is chosen such that the normalization of the observed $L_X-E_{\gamma\rm ,iso}$ correlation is reproduced (see Fig. \ref{fig:normalization}).
For each simulation, the left panel shows the simulated $E_{\gamma\rm ,iso}-L_X$ relation for 43 randomly selected bursts (circles denote bursts with $\nu_c<2$keV, 
pluses, bursts with $2\,\rm keV<\nu_c<10\,$keV and X's, bursts with $\nu_c>10\,$keV). 
Grey crosses refer instead to the 43 GRBs in the sample of \citealt{D'Avanzo(2012)} (see also Fig. \ref{fig:observations}).
Solid lines depict the best linear fits and dashed lines depict the $3 \sigma$ scatter of the simulated correlation. The panel on the right shows the ratio between the kinetic energies derived using Eq.~\ref{eq:standard} and the simulated (see text for details).}
\label{fig:epsB}
\end{figure*}

Since the main parameter determining the results is $\epsilon_B$, in Fig. \ref{fig:epsB} we show the resulting $L_{X}-E_{\gamma\rm ,iso}$ correlation for 43 simulated bursts (so as to fit the number of bursts in the BAT6 sample)
for three cases: $\langle \log_{10}\epsilon_B\rangle=-2$ (upper left panel), $\langle \log_{10}\epsilon_B\rangle=-4$ (middle left panel) and $\langle \log_{10}\epsilon_B\rangle=-6$ (lower left panel).
For each simulation, Fig. \ref{fig:epsB} also shows the ratio between the kinetic energies inferred 
using eq. \ref{eq:standard} and the simulated one (panels on the right). 
For $\langle \log_{10}\epsilon_B\rangle =-4$ ($\langle \log_{10}\epsilon_B\rangle =-2$), we get $E_{\rm kin,2}/E_{\rm kin,real}=0.04_{-0.03}^{+0.08}$
($E_{\rm kin,2}/E_{\rm kin,real} =0.28_{-0.17}^{+0.45}$).
Naturally, this affects also the estimates of the prompt efficiencies.
In Fig. \ref{fig:efficiencyratio} we explicitly show how the ratio of the derived to real efficiency varies as a function of the mean value of $\epsilon_B$ for both ISM and wind environments.
In both cases $\epsilon_B \lesssim 10^{-3}$ leads to a significant deviation (of order $\gtrsim 2$) of the derived efficiency as compared with the real one.
\begin{figure*}
\hskip 0.2truecm
\includegraphics[scale=0.3]{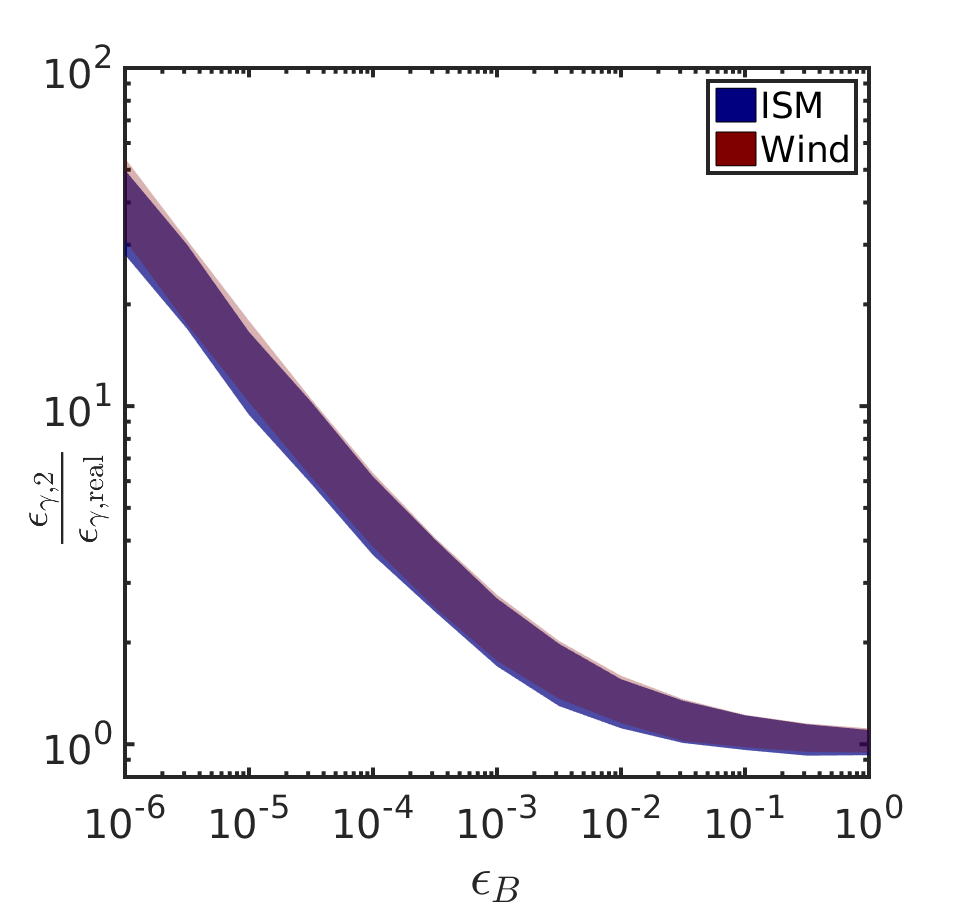}
\caption{The ratio of the derived prompt $\gamma$-ray efficiency (assuming fast cooling synchrotron with no IC suppression) compared to the actual simulated efficiency as a function of $\epsilon_B$ for an ISM and a wind.
In all simulations we use $\langle\log_{10}(n [\mbox{cm}^{-3}])\rangle=0$ ($\langle \log_{10}(A_*)\rangle=0$), and standard deviations of 1dex in the density (or wind parameter) and $\epsilon_B$ and $p=2.2$.}
\label{fig:efficiencyratio}
\end{figure*}

\section{Discussion and Conclusions}\label{sec:conclusions}

The kinetic energy of the blast wave (during the afterglow phase) and the corresponding efficiency of the prompt phase are among the most important parameters concerning the emission regions in GRBs.
Following \cite{Kumar(2000)} and \cite{Freedman(2001)} the \xr luminosity at 11 hours has been traditionally used to infer the kinetic energy \citep{Berger(2003), LRZ(2004), Berger(2007),Nysewander(2009),D'Avanzo(2012),Wygoda(2015)}
resulting usually in very large prompt efficiencies. This method has been claimed to be quite robust, since no other quantities apart from $\epsilon_e $ and $\epsilon_{\gamma}$ are involved. 
We have re-investigated the question whether the \xr are indeed a good proxy for $E_{\rm kin}$. 
This is motivated by the recent findings that the typical values of $\epsilon_B$ might be much smaller than the values 0.01-0.1
traditionally assumed. An additional line of motivation is the apparent contradiction between energies estimated in this way using the \xr flux as compared with the energies estimated using the 0.1-$10\,$GeV radiation detected by Fermi/LAT
\citep{Beniamini(2015)}. In that paper we have argued that this contradiction can be resolved within the synchrotron forward shock scenario if the X-ray emitting electrons are either slow cooling or else, they are
strongly affected by IC cooling: in both cases, the X-ray emission is not a good proxy for the energy of the blast wave.
These conclusions are however model dependent, since they rely on the assumption that the GeV radiation is synchrotron emission from the external
shock.
Other studies considered an alternative possibility, in which the GeV radiation is not of afterglow origin \citep[e.g.][]{Beloborodov(2014)}.

For $\epsilon_B\sim 0.01-0.1$ (and $\epsilon_e\sim0.1$), SSC losses are small and the afterglow synchrotron luminosity above the characteristic synchrotron frequencies is proportional to $\epsilon_e$
times the kinetic energy of the blast wave $E_{\rm kin}$.
The relation between the \xr flux and the kinetic energy (eq. \ref{eq:standard}) depends very weakly on $\epsilon_B$ and is independent of the density. 
Thus, the observed correlation between $L_X$ and $E_{\gamma\rm ,iso}$ gave support to the fact that $L_X$ can be used to infer $E_{\rm kin}$.

For smaller values of $\epsilon_B$, SSC cannot be ignored and $L_X$ depends indirectly on $\epsilon_B$.
This is the main parameter regulating the importance of SSC vs. synchrotron emission as well as determining whether X-ray emitting electrons are slow or fast cooling.  
We show here that somewhat surprisingly the observed $L_X-E_{\gamma\rm ,iso}$ correlation 
is recovered also when the full effect of $\epsilon_B$ on $L_X$ is taken into account. For small $\epsilon_B$ values, $L_{\rm GeV}$ (not affected by SSC cooling) rather than 
$L_X$, is a good proxy for the kinetic energy, and is indeed strongly correlated with $E_{\gamma \rm,iso}$ \citep{Nava(2014)}.

SSC cooling modifies the synchrotron spectrum so that the cooling frequency is $\nu_c=\nu_c^{syn}/(1+Y)^2$ and the luminosity above $\nu_c$ is $L(\nu>\nu_c)=L(\nu>\nu_c)^{syn}/(1+Y)$.
By means of analytic and numerical estimates we found that the X-ray frequency most likely lies in this part of the synchrotron spectrum, even for small $\epsilon_B\sim 10^{-6}$.
The observed $L_X$ is then suppressed by a factor $(1+Y)$. 
This factor, $(1+Y)$, depends only weakly on the energy and the relation between $L_X$ and $E_{\gamma\rm ,iso}$ is still linear. 
This means that  approximately $L_X/E_{\gamma\rm ,iso} \propto \epsilon_e^{1/2}  \epsilon_B^{1/2} \epsilon_\gamma^{-1} $ (with possibly a weak dependence on $n$ for small $n$ values),
instead of  $\propto \epsilon_e \epsilon_\gamma^{-1}$.
While an additional parameter was added, the dependence on both it and $\epsilon_e$ is smaller than before, and hence it is reasonable to have a comparable spread. 
The observed correlation is reproduced under very reasonable assumptions (Table~\ref{tbl:dispersions}). 
The normalization and scatter of the correlation can be recovered even with very small values of $\epsilon_B\gtrsim 10^{-6}$, demonstrating that the recent findings of small
magnetic field are not at odds with the existence of a clear trend between $L_X$ and $E_{\gamma\rm ,iso}$.

We reconfirm the results of our previous work \citep{Beniamini(2015)}, that generally,   
$L_X$ is not a good proxy for the kinetic energy and that on its own the GeV afterglow luminosity, $L_{GeV}$ is much better proxy for the blast wave kinetic energy. When both are combined,
both this energy and $\epsilon_B$ can be determined.
Including IC corrections to $L_X$, we find larger kinetic energies and lower efficiencies than reported in studies assuming no IC suppression.
More specifically, lower values of the prompt efficiency ($\epsilon_\gamma\lesssim0.2$), can be accounted for by invoking lower values of the magnetic field ($\epsilon_B\lesssim10^{-4}$),
while if larger values of $\epsilon_B$ are assumed, then larger values of the prompt efficiency must be invoked to match the observations.
Estimates of the kinetic blast wave energies are fundamental not only to determine the energetics of the system, but also to infer the efficiency of the mechanism producing the prompt radiation
(i.e. the ratio between the energy radiated in the prompt phase $E_{\gamma\rm ,iso}$ and the initial outflow energy $E_{\gamma\rm ,iso}+E_{\rm kin}$). 
In the past, the large inferred value of $\epsilon_{\gamma}$ has been claimed as one of the main arguments against the internal shock model, within which large efficiencies can hardly be achieved.
In fact, obtaining order unity efficiencies is very problematic in a wide range of models, including most models invoking magnetic reconnection.
Thus, reducing the requirements on the efficiency, opens up somewhat the parameter space of allowed prompt models and we may have to reconsider our picture of the prompt phase in light of these results.

We thank Rodolfo Barniol Duran and Pawan Kumar for helpful comments.
We acknowledge support by the John Templeton Foundation, by  a grant from the Israel Space Agency, by a ISF-CNSF grant and by the ISF-CHE I-core center
for excellence for research in Astrophysics. LN was supported by a Marie Curie Intra-European Fellowship of the European Community's 7th Framework Programme (PIEF-GA-2013- 627715).


\end{document}